\documentclass{webofc}
\usepackage[varg]{txfonts}

\usepackage{color}
\usepackage[colorlinks,linkcolor=blue,citecolor=blue,urlcolor=blue]{hyperref}

\usepackage{xspace}
\newcommand{\sNN}{\ensuremath{\sqrt{s_\mathrm{NN}}}\xspace}

\newcommand{\AuAu}{Au~+~Au\xspace}

\begin{document}

\title{Global hyperon polarization and effects of decay feeddown}

\author{%
    \firstname{Hui} \lastname{Li}\inst{1}\fnsep\thanks{\email{lihui_fd@fudan.edu.cn}} \and
    \firstname{Xiao-Liang} \lastname{Xia}\inst{2}\fnsep\thanks{\email{xiaxl@fudan.edu.cn}} \and
    \firstname{Xu-Guang} \lastname{Huang}\inst{1,2}\fnsep\thanks{\email{huangxuguang@fudan.edu.cn}} \and
    \firstname{Huan Zhong} \lastname{Huang}\inst{1,3}\fnsep\thanks{\email{huanzhonghuang@fudan.edu.cn}}
}

\institute{%
    Key Laboratory of Nuclear Physics and Ion-beam Application (MOE), Fudan University, Shanghai 200433, China
    \and
    Department of Physics and Center for Field Theory and Particle Physics, Fudan University, Shanghai 200433, China
    \and
    Department of Physics and Astronomy, University of California, Los Angeles, CA 90095, USA
}

\abstract{%
    We study the global polarizations of $\Lambda$, $\Xi^-$, and $\Omega^-$ hyperons in noncentral \AuAu collisions at $\sqrt{s_\mathrm{NN}}=$ 7.7--200 GeV. We highlight the importance of effect of decay feeddown to the measured global polarization. With the decay contributions taken into account, the global polarization ordering $P_{\Omega^-} > P_{\Xi^-} > P_\Lambda$ can be naturally explained, which is consistent with the observation recently reported from the STAR experiment from Au+Au collisions at 200 GeV. We also extend our calculations to predict expectations from the RHIC-BES II data.
}

\maketitle

\section{Introduction}

The large orbital angular momentum produced by relativistic heavy-ion collisions leads to the spin polarization of final-state hadrons~\cite{Liang:2004ph,Voloshin:2004ha}. Therefore, the spin polarization in turn provides us with a tool to detect the rotational motion of the quark-gluon plasma (QGP) created in the collision. The global polarization of $\Lambda$ and $\bar\Lambda$ hyperons, which reflects the average vorticity strength of the QGP, has been measured by the STAR~\cite{STAR:2017ckg,Adam:2018ivw,STAR:2021beb}, the ALICE~\cite{Acharya:2019ryw}, and the HADES~\cite{HADES2021} experiments in a wide collision energy range covering $\sNN=$ 2.4--5020 GeV. The data in the energy range of $\sNN=$ 7.7--5020 GeV~\cite{STAR:2017ckg,Adam:2018ivw,Acharya:2019ryw} demonstrate an increasing global $\Lambda$ ($\bar\Lambda$) polarization with the decreasing collision energy. This trend can be reproduced by many model calculations, such as~\cite{Karpenko:2016jyx,Li:2017slc,Xie:2017upb,Shi:2017wpk,Wei:2018zfb}, which provides strong support to the vorticity interpretation of the global polarization.

In order to further examine the global polarization mechanism, it is important to study the spin polarization (or spin alignment) of other hadron species. Recently, the STAR experiment has measured the global polarizations of $\Xi$ and $\Omega$ hyperons at $\sNN=$ 200 GeV~\cite{Adam:2020pti}. The STAR and the ALICE experiments have also measured the spin alignment of $\phi$ and $K^{*0}$ mesons~\cite{Zhou:2019lun,Acharya:2019vpe,Singha:2020qns}. However, the observed deviation of the spin-density matrix element $\rho_{00}$ from 1/3 does not necessarily indicate the global polarization. Local polarization distribution caused by the QGP anisotropic expansion, which does not contribute to the global polarization, can also lead to $\rho_{00}\neq 1/3$~\cite{Xia:2020tyd}. This makes the spin alignment of vector mesons more complicated than the global polarization of hyperons.

In this proceeding, we report our recent research~\cite{Li:2021zwq} on the global polarizations of $\Lambda$, $\Xi^-$, and $\Omega^-$ hyperons within the framework based on thermal vorticity and spin transfer dynamics in decay. The study improves our understanding of the global hyperon polarization and provides predictions for future measurements at RHIC-BES energies.

\section{Framework}

We calculated the global polarizations of $\Lambda$, $\Xi^-$, and $\Omega^-$ hyperons in \AuAu collisions at RHIC energies $\sNN=$ 7.7--200 GeV. The calculation can be divided into the following two procedures.

We first compute the global polarization of \textit{primary} $\Lambda$, $\Xi^-$, and $\Omega^-$ hyperons. The primary hyperons are those produced by hadronization at the end of QGP fluid. Spin statistical and kinetic theory shows that their spin polarizations are determined by thermal vorticity of the fluid~\cite{Becattini:2013fla,Fang:2016vpj,Liu:2020flb}:
\begin{equation}
    P^\mu\simeq -\frac{S+1}{6m}\epsilon^{\mu\nu\rho\sigma}p_\nu\varpi_{\rho\sigma}(x),
    \label{eq-PH}
\end{equation}
where $P^\mu$ is the spin polarization four-vector, and $S$, $m$, $p$, and $x$ are the spin, mass, momentum four-vector, and space-time coordinate four-vector of the hyperon. The thermal vorticity is defined as $\varpi_{\rho\sigma}=\tfrac{1}{2}(\partial_\sigma\beta_\rho-\partial_\rho\beta_\sigma)$ with $\beta_\rho=u_\rho/T$ in which $u$ is the fluid four-velocity and $T$ is the temperature. We extract the thermal vorticity field and the phase-space distributions of primary $\Lambda$, $\Xi^-$, and $\Omega^-$ hyperons from a multi-phase transport (AMPT) model~\cite{Lin:2004en}. With this information, the global polarization of primary hyperons can be calculated using Eq.~(\ref{eq-PH}). More details about the calculation can be found in~\cite{Li:2021zwq}.

Then we estimate the decay contributions to the global polarization of \textit{inclusive} hyperons. Note that the global polarization calculated in the first procedure is of the primary hyperons. However, the experimental measurements include the hyperons produced by decay of heavier particles. Through a statistical model calculation, we find that about 80\% of the final $\Lambda$’s and about 40\% of the final $\Xi^-$'s are products of decays. Following from Eq.~(\ref{eq-PH}), the thermal vorticity polarizes not only the primary $\Lambda$, $\Xi^-$, and $\Omega^-$ hyperons, but also all other hadron species created from the QGP (except spinless hadrons). When these polarized hadrons decay, their spin polarizations are transferred to secondary particles. The detailed spin-transfer rules can be found in~\cite{Li:2021zwq,Becattini:2016gvu,Becattini:2019ntv,Xia:2019fjf}. Taking the decay contributions into account, the global polarization of inclusive hyperons can be calculated by
\begin{equation}
    P_{H,\text{inclusive}} = \frac{N_H P_{H,\text{primary}} + \sum N_{X\to H} P_{X\to H}} {N_H + \sum N_{X\to H}},
    \label{eq-inclusive}
\end{equation}
where $H$ is the hyperon which we want to study, $X$ stands for particles which can decay to $H$, $P_{H,\text{inclusive}}$ ($P_{H,\text{primary}}$) is the global polarization of inclusive (primary) $H$ hyperon, $N_H$ is the primary yield number of $H$, $N_{X\to H}$ and $P_{X\to H}$ are the yield number and the global polarization of $H$ from the decay $X\!\to\!H$, and the sum runs over all possible decays (including cascade decays) from $X$ to $H$. We use the THERMUS package~\cite{Wheaton:2004qb} (which is a statistical model) to calculate the primary yields of the following hyperons: $\Lambda$, $\Lambda(1405)$, $\Lambda(1520)$, $\Lambda(1600)$, $\Lambda(1670)$, $\Lambda(1690)$, $\Sigma^0$, $\Sigma(1385)$, $\Sigma(1660)$, $\Sigma(1670)$, $\Xi$, and $\Xi(1530)$. We then use the spin-transfer rules to simulate decay of these hyperons and obtain $N_{X\to H}$ and $P_{X\to H}$. Finally, these quantities are substituted into Eq.~(\ref{eq-inclusive}) to calculate the inclusive global polarization.

\section{Results and discussion}

\begin{figure}
    \centering
    \includegraphics[width=0.5\textwidth]{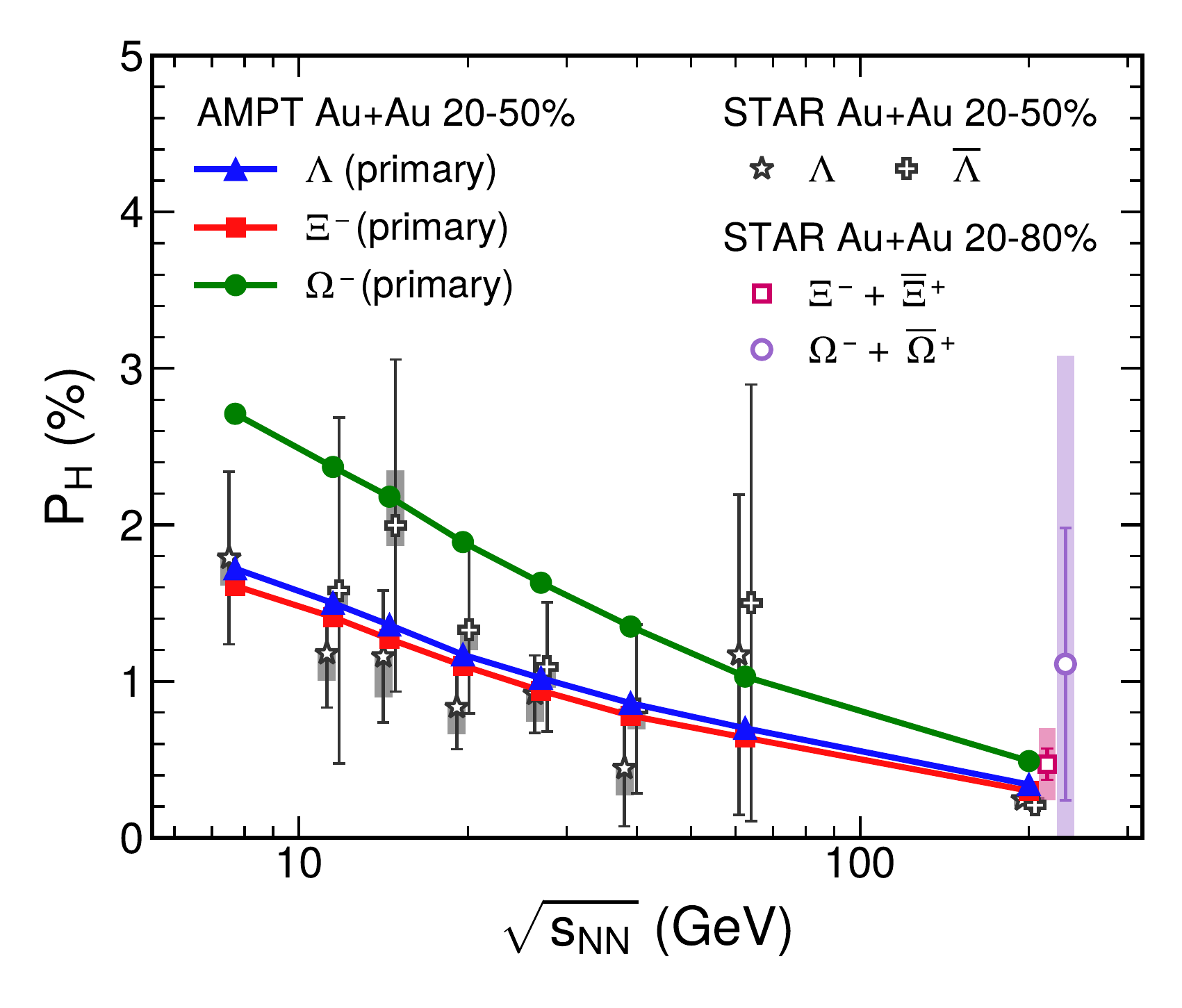}%
    \includegraphics[width=0.5\textwidth]{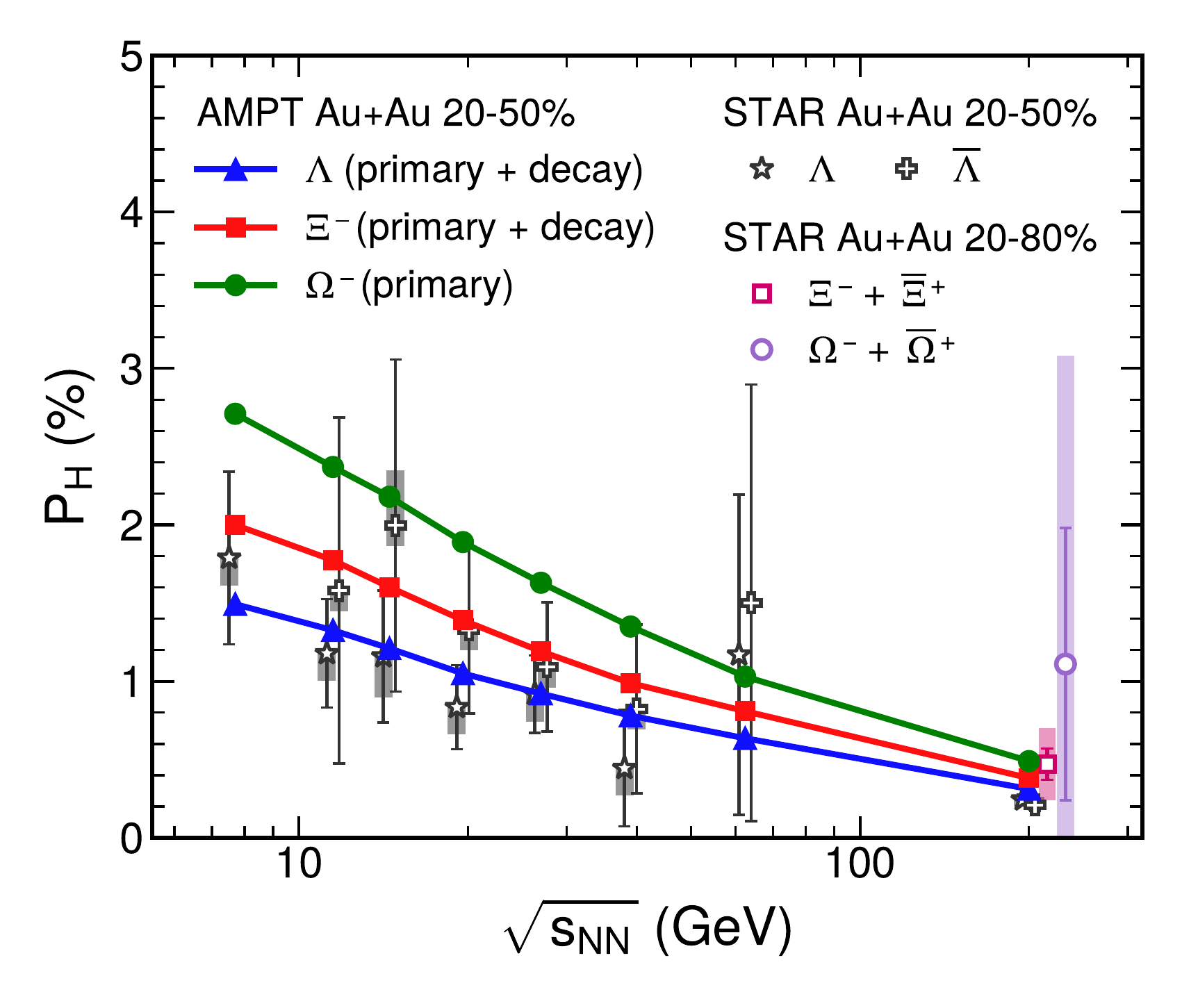}
    \caption{Global polarizations of $\Lambda$, $\Xi^-$, and $\Omega^-$ hyperons without (left) and with (right) the decay contributions to $\Lambda$ and $\Xi^-$ taken into account. The STAR data~\cite{STAR:2017ckg,Adam:2018ivw,Adam:2020pti} of inclusive hyperon polarizations are also shown for comparison.}
    \label{fig-PH}
\end{figure}

In this section, we present and discuss our results on the global hyperon polarizations. In the current framework, we have assumed that the primary hyperons follow the same global polarization mechanism, that is, they are polarized by the thermal vorticity through Eq.~(\ref{eq-PH}). Therefore, we expect that the global polarizations of the three hyperons have a similar energy dependence --- they all decrease with the increase of collision energy, as demonstrated in Fig.~\ref{fig-PH}.

The left panel of Fig.~\ref{fig-PH} shows that the global polarizations of primary $\Lambda$, $\Xi^-$, and $\Omega^-$ hyperons approximately fulfill $P_\Lambda:P_{\Xi^-}:P_{\Omega^-}\simeq 1:1:5/3$. The main reason is that $\Omega^-$ is a spin-3/2 hyperon, while $\Lambda$ and $\Xi^-$ are spin-1/2, and equation (\ref{eq-PH}) shows that the global polarization is proportional to $S+1$. In addition to the spin number, equation~(\ref{eq-PH}) also indicates that the global polarization depends on hyperon's four-velocity $(p_\nu/m)$ and space-time coordinate $x$. However, as we examined in the AMPT model, the three hyperons have very similar velocity and space-time distributions. As a result, the difference between the primary global polarizations of $\Lambda$ and $\Xi^-$ is very small. Recent hydrodynamic calculation~\cite{Fu:2020oxj} yielded similar conclusion.

The right panel of Fig.~\ref{fig-PH} shows the global polarizations of $\Lambda$, $\Xi^-$, and $\Omega^-$ hyperons, but with the decay contributions to $\Lambda$ and $\Xi^-$ taken into account. Compared with the primary global polarizations (Fig.~\ref{fig-PH} left), decay of heavier particles reduces the global polarization of $\Lambda$ by about 10\%, but increases the global polarization of $\Xi^-$ by about 25\%~\cite{Li:2021zwq}. The $\Omega^-$ hyperon contains hardly any contribution from decays, therefore we neglect the decay contributions to the global $\Omega^-$ polarization. Finally, after we consider the decay contributions, the global polarizations of inclusive $\Lambda$ and $\Xi^-$ hyperons become separated, and we find $P_{\Omega^-}>P_{\Xi^-}>P_\Lambda$. This explains the recent experimental measurement at 200 GeV~\cite{Adam:2020pti}. Our calculations predict that the ordering of $P_{\Omega^-}>P_{\Xi^-}>P_\Lambda$ will become more prominent from the RHIC-BES II data if there would be sufficient statistics to make the measurements.

\section{Summary}

In summary, we studied the global polarizations of $\Lambda$, $\Xi^-$ and $\Omega^-$ hyperons in \AuAu collisions at RHIC energies $\sNN=$ 7.7--200 GeV, within the framework based on the thermal vorticity and spin transfer in decay. We find that the global polarizations of primary hyperons approximately fulfill $P_\Lambda:P_{\Xi^-}:P_{\Omega^-}\simeq 1:1:5/3$, and the global polarizations of inclusive hyperons are in the ordering: $P_{\Omega^-}>P_{\Xi^-}>P_\Lambda$. We have analyzed multiple reasons for the relations among $P_\Lambda$, $P_{\Xi^-}$, and $P_{\Omega^-}$. We found that $\Omega^-$ has the largest global polarization, which is due to its larger spin number ($S=3/2$). On the other hand, the decay contributions play an important role in the separation between the inclusive global polarizations of $\Lambda$ and $\Xi^-$. Our findings can naturally explain the recent experimental data at 200 GeV~\cite{Adam:2020pti}. We also expect more experimental measurements at RHIC-BES energies to further test our results.

\section*{Acknowledgment}

This work is supported in part by NSFC under grants Nos.~11835002 and 12075061, by Shanghai Natural Science Foundation under grant No.~20ZR1404100, and by China Postdoctoral Science Foundation under grant Nos.~2018M641909 and 2019M661333.


\begin{thebibliography}{}
    \bibitem{Liang:2004ph}
    Z.~T.~Liang and X.~N.~Wang,
    Phys. Rev. Lett. \textbf{94}, 102301 (2005)

    \bibitem{Voloshin:2004ha}
    S.~A.~Voloshin,
    arXiv:nucl-th/0410089 [nucl-th]

    \bibitem{STAR:2017ckg}
    L.~Adamczyk \textit{et al.} [STAR],
    Nature \textbf{548}, 62-65 (2017)

    \bibitem{Adam:2018ivw}
    J.~Adam \textit{et al.} [STAR],
    Phys. Rev. C \textbf{98}, 014910 (2018)

    \bibitem{STAR:2021beb}
    M.~S.~Abdallah \textit{et al.} [STAR],
    arXiv:2108.00044 [nucl-ex]

    \bibitem{Acharya:2019ryw}
    S.~Acharya \textit{et al.} [ALICE],
    Phys. Rev. C \textbf{101}, no.4, 044611 (2020)

    \bibitem{HADES2021}
    F.~Kornas [HADES],
    these proceedings (2021)

    \bibitem{Karpenko:2016jyx}
    I.~Karpenko and F.~Becattini,
    Eur. Phys. J. C \textbf{77}, no.4, 213 (2017)

    \bibitem{Li:2017slc}
    H.~Li, L.~G.~Pang, Q.~Wang and X.~L.~Xia,
    Phys. Rev. C \textbf{96}, no.5, 054908 (2017)

    \bibitem{Xie:2017upb}
    Y.~Xie, D.~Wang and L.~P.~Csernai,
    Phys. Rev. C \textbf{95}, no.3, 031901 (2017)

    \bibitem{Shi:2017wpk}
    S.~Shi, K.~Li and J.~Liao,
    Phys. Lett. B \textbf{788}, 409-413 (2019)

    \bibitem{Wei:2018zfb}
    D.~X.~Wei, W.~T.~Deng and X.~G.~Huang,
    Phys. Rev. C \textbf{99}, no.1, 014905 (2019)

    \bibitem{Adam:2020pti}
    J.~Adam \textit{et al.} [STAR],
    Phys. Rev. Lett. \textbf{126}, no.16, 162301 (2021)

    \bibitem{Zhou:2019lun}
    C.~Zhou [STAR],
    Nucl. Phys. A \textbf{982}, 559-562 (2019)

    \bibitem{Acharya:2019vpe}
    S.~Acharya \textit{et al.} [ALICE],
    Phys. Rev. Lett. \textbf{125}, no.1, 012301 (2020)

    \bibitem{Singha:2020qns}
    S.~Singha [STAR],
    Nucl. Phys. A \textbf{1005}, 121733 (2021)

    \bibitem{Xia:2020tyd}
    X.~L.~Xia, H.~Li, X.~G.~Huang and H.~Z.~Huang,
    Phys. Lett. B \textbf{817}, 136325 (2021)

    \bibitem{Li:2021zwq}
    H.~Li, X.~L.~Xia, X.~G.~Huang and H.~Z.~Huang,
    arXiv:2106.09443 [nucl-th]

    \bibitem{Becattini:2013fla}
    F.~Becattini, V.~Chandra, L.~Del Zanna and E.~Grossi,
    Annals Phys. \textbf{338}, 32-49 (2013)

    \bibitem{Fang:2016vpj}
    R.~H.~Fang, L.~G.~Pang, Q.~Wang and X.~N.~Wang,
    Phys. Rev. C \textbf{94}, no.2, 024904 (2016)

    \bibitem{Liu:2020flb}
    Y.~C.~Liu, K.~Mameda and X.~G.~Huang,
    Chin. Phys. C \textbf{44}, no.9, 094101 (2020)

    \bibitem{Lin:2004en}
    Z.~W.~Lin, C.~M.~Ko, B.~A.~Li, B.~Zhang and S.~Pal,
    Phys. Rev. C \textbf{72}, 064901 (2005)

    \bibitem{Becattini:2016gvu}
    F.~Becattini, I.~Karpenko, M.~Lisa, I.~Upsal and S.~Voloshin,
    Phys. Rev. C \textbf{95}, no.5, 054902 (2017)

    \bibitem{Becattini:2019ntv}
    F.~Becattini, G.~Cao and E.~Speranza,
    Eur. Phys. J. C \textbf{79}, no.9, 741 (2019)

    \bibitem{Xia:2019fjf}
    X.~L.~Xia, H.~Li, X.~G.~Huang and H.~Z.~Huang,
    Phys. Rev. C \textbf{100}, no.1, 014913 (2019)

    \bibitem{Wheaton:2004qb}
    S.~Wheaton and J.~Cleymans,
    Comput. Phys. Commun. \textbf{180}, 84-106 (2009)

    \bibitem{Fu:2020oxj}
    B.~Fu, K.~Xu, X.~G.~Huang and H.~Song,
    Phys. Rev. C \textbf{103}, no.2, 024903 (2021)
\end{thebibliography}
\end{document}